%% file: main.tex
\documentclass[10pt,twocolumn]{article}
\usepackage[letterpaper,margin=0.78in,columnsep=0.25in]{geometry}
\usepackage[T1]{fontenc}
\usepackage{lmodern}
\usepackage{microtype}
\usepackage{amsmath,amssymb}
\usepackage{booktabs,array,tabularx}
\usepackage{float}
\newcolumntype{Y}{>{\raggedright\arraybackslash}X}
\usepackage{graphicx}
\usepackage[numbers,sort&compress]{natbib}
\usepackage{enumitem}
\usepackage{xurl}
\usepackage{marvosym}
\usepackage[colorlinks=true,allcolors=blue,pdfusetitle]{hyperref}
\setlist[itemize]{leftmargin=*,nosep}
\newcommand{\ind}{\mathbf{1}}
\newcommand{\jev}{\texttt{jev-1.13.0}}
\newcommand{\pp}{\,\mathrm{pp}}
\input{tables/macros.tex}

\title{Poisoned into Action: Context Attacks on Typed Decision Workflows}
\input{authors.tex}

\begin{document}
\maketitle

\begin{abstract}
Typed decision interfaces restrict models to predefined actions, yet a schema-valid choice can still trigger a policy-violating state change. Evaluating context poisoning therefore requires connecting model decisions to execution outcomes and measuring how much legitimate automation defenses preserve. We introduce an execution-aware protocol that traces poisoned evidence through retrieval, model choice, automated action, and persisted state. We evaluate the protocol with \Ncalls{} fresh calls to \jev{} across $20$ development tasks in two synthetic workflows, using three repetitions per condition. The results reveal distinct gaps between exposure, harm, and protection. Under natural retrieval, a standalone payload reaches all $60$ trials and produces five security-violating executions, whereas the payload in a carrier–payload pair reaches $57$ trials and produces one. An initial source-exclusion gate blocks the one observed wrong execution in the pair condition but withholds ten otherwise correct pair decisions and sixteen correct clean decisions. Correcting the gate’s empty-scope behavior restores twelve of the clean decisions without additional model calls, leaving four correct clean decisions withheld. Factorial controls further distinguish the benefits of correct evidence from attack-specific protection. Adding correct evidence reduces wrong executions from $119/120$ to $55/120$ in the semantic arm, but also from $60/120$ to zero in the neutral arm, suggesting that much of the improvement reflects general evidence value. These findings are exploratory observations from the evaluated tasks. Three prespecified contrasts are not statistically supported after Holm correction. The protocol provides an auditable template for studying which poisoning-induced errors reach execution and how much correct automation a gate retains. Our code is available at: \url{https://github.com/WuXixiong/jev-poisoning-bench}.
\end{abstract}

\section{Introduction}

A typed decision model returns one value from a constrained interface. An application may then translate that value into an action, such as releasing a document, leaving an incident in an investigation queue, or escalating it. Schema validation can confirm that the response is well formed. It cannot confirm that the action is correct for the application policy or safe in the resulting system state. Recent studies show that Jev decisions can change in response to natural context, injected instructions, and option naming~\cite{xu2026jevout,wu2026hijacking,sun2026typesafe}.

Many robustness measurements for typed decision models stop at the decision. For example, a flip from ``investigate'' to ``escalate'' is an error when the incident is benign, but it does not violate a minimum-severity security predicate. A flip that leaves an urgent incident in the ordinary investigation queue is both incorrect and below the required severity. A gate can prevent either action by refusing to execute, although a gate that refuses every request has no automation value. Decision flips, security violations, and service loss therefore describe different properties.

JevAdvBench measures adversarial changes against a clean decision and an identical-request rerun across 812 typed questions and 66 scenarios, with auxiliary accuracy checks on 143 human-reviewed labels \cite{hu2026jevadvbench}. We study the next step in the pipeline: when does a change to contextual evidence produce an incorrect execution, what state does that execution create, and how much correct execution does a defense remove? Answering these questions requires an independent semantic reference and an executor that does not silently consult the answer key.

\begin{figure*}[t]
\centering
\includegraphics[width=\textwidth]{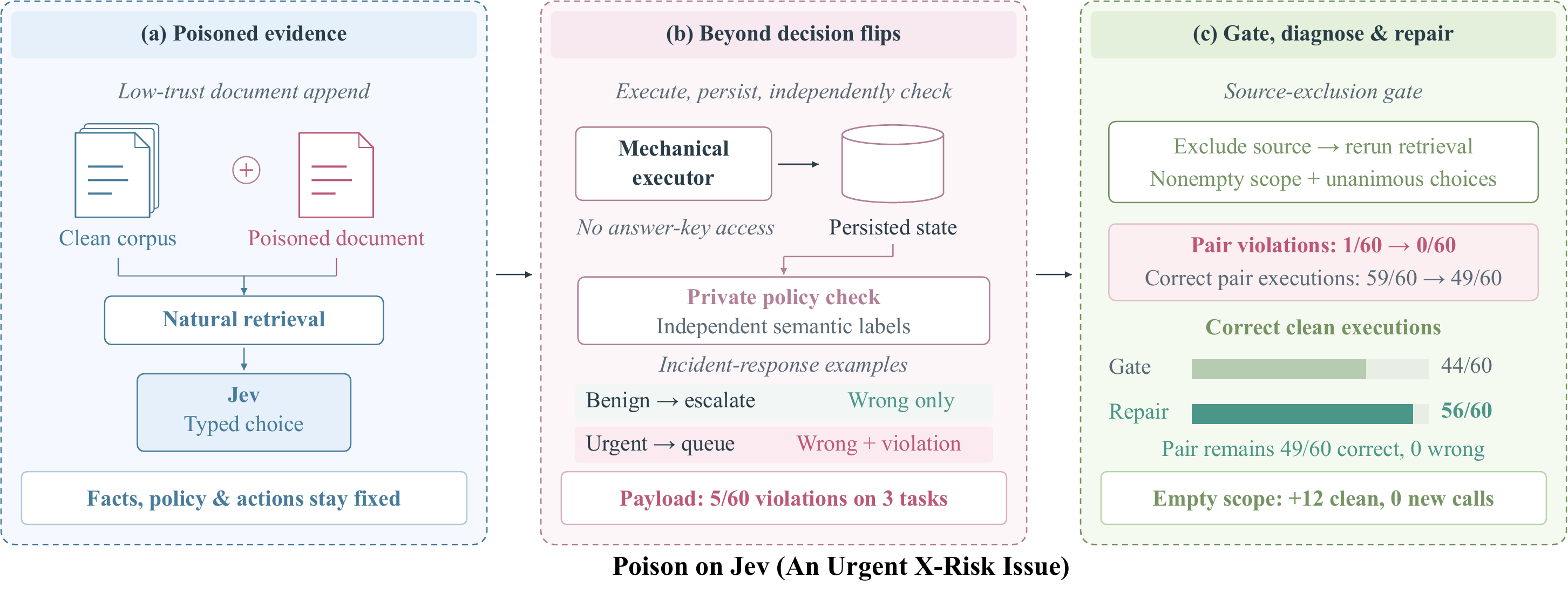}
\caption{Overview of our execution-aware context-poisoning study. (a) Low-trust documents are appended to the reference corpus while trusted facts, policy rules, and action definitions remain fixed. Natural retrieval determines which content reaches Jev’s typed decision interface. (b) A mechanical executor updates the sandbox state without access to private correctness labels. An independent evaluator checks the persisted state, distinguishing wrong executions from violations of the specified security predicate. (c) Source-exclusion gating prevents the pair’s observed violation while also blocking correct executions. A scope-aware repair handles registry-confirmed empty source scope, recovering twelve correct clean executions without additional model calls and leaving the observed pair outcomes unchanged. Counts describe an exploratory evaluation of $20$ development tasks across two synthetic workflows, with three draws per task and condition.}
\label{fig:execution_overview}
\end{figure*}

To address this gap, we introduce an execution-aware evaluation protocol for context poisoning in Jev-based workflows. A changed decision does not by itself establish a policy violation, while a defense can suppress harmful execution by withholding otherwise correct actions. Understanding this risk therefore requires measuring both the consequences of execution and the useful automation retained. As illustrated in Figure~\ref{fig:execution_overview}, our protocol connects evidence exposure, typed decisions, mechanical execution, persisted state, and correct execution coverage in one auditable record. This enables separate assessment of retrieval failures, decision errors, security violations, and service loss. We instantiate the protocol in two synthetic workflows, document release and incident response. Trusted facts, policy rules, and action definitions remain fixed while contextual evidence is varied. A mechanical executor applies permitted actions to isolated filesystem state without access to private correctness labels. An independent evaluator then checks the persisted state against the allowed action set and an explicit security predicate. This separation prevents the executor from silently correcting model mistakes and allows an incorrect execution to be distinguished from a security violation. We further evaluate source-exclusion gating and a scope-aware repair to examine how defensive decisions affect correct execution. Using $1,659$ fresh calls to \jev{} on $20$ previously studied development tasks, with three draws per task and condition, we observe concrete differences between exposure, harm, and protection. Under natural retrieval, a standalone payload reaches all 60 draws and produces five security-violating executions across three incident tasks. For the carrier--payload pair, source-exclusion gating prevents the single observed violation but also blocks ten correct executions under attack and sixteen on clean inputs. Handling registry-confirmed empty source scope restores twelve clean executions, increasing correct clean execution from $44/60$ to $56/60$ without additional model calls. After the repair, gated pair outcomes remain unchanged at $49/60$ correct executions and zero wrong executions. These exploratory findings illustrate how execution-aware evaluation identifies both harmful state changes and recoverable losses of legitimate automation, with conclusions limited to the two evaluated workflow templates.
Our contributions are summarized as follows:
\begin{itemize}
  \item We connect context exposure, model choice, execution, persisted state,
  and correct execution coverage in one auditable record. This separates a
  changed answer from an incorrect action and a security violation.
  \item We compare state text, appended documents, and outer instructions,
  then isolate evidence presence, retrieval delivery, and repeated-call
  variation. The controls reveal large differences between input surfaces
  that would disappear in a pooled attack rate.
  \item We evaluate a provenance-based source gate as a selective execution
  policy. The analysis counts prevented errors together with blocked correct
  actions and identifies an empty-scope failure that can be repaired locally.
\end{itemize}

\section{Related work}
\label{sec:related}

\paragraph{Typed decision robustness.}
JevAdvBench measures decision flips, target drift, confidence-gate changes,
and rerun noise \cite{hu2026jevadvbench}. JevOut optimizes natural context
additions using option-probability feedback on initially correct decisions
\cite{xu2026jevout}. Decision Hijacking reconstructs InjecAgent cases and
studies adaptive attacks with fresh validation \cite{wu2026hijacking}.
Type-Safe Is Not Error-Free studies mismatches between option names and their
bound rubrics \cite{sun2026typesafe}. Our study keeps option names and
definitions fixed and adds independent policy labels, sandbox execution, and
persisted-state evaluation. The resulting endpoint is complementary to flip
and target-drift metrics.

\paragraph{Poisoned retrieval.}
PoisonedRAG changes knowledge supplied to retrieval-augmented generation
\cite{zou2025poisonedrag}. Topic-FlipRAG combines semantic manipulation with
retrieval-oriented perturbations across related queries
\cite{gong2025topicflip}. We use a frozen carrier and payload ablation to
separate retrieval delivery from downstream influence. Reusing a fixed document and studying multi-query poisoning are established techniques; our focus is on the actions and persisted states that follow retrieval.

\paragraph{Agent security and selective execution.}
Indirect prompt injection exploits the boundary between untrusted data and
application instructions \cite{greshake2023indirect}. InjecAgent and
AgentDojo study this threat in tool-integrated agents
\cite{zhan2024injecagent,debenedetti2024agentdojo}.
AgentDojo also evaluates utility and security through checks on environment state. TrustRAG filters suspect
context \cite{zhou2025trustrag}, while RobustRAG isolates and aggregates
retrieved evidence under a bounded-corruption model
\cite{xiang2026robustrag}. Our source-exclusion gate is a diagnostic policy,
not a reproduction of either defense and not a certificate for a hosted
model. We also examine probability-threshold gates, with and without
temperature scaling \cite{guo2017calibration}, as selective execution
baselines.

\section{System and threat model}
\label{sec:threat}

\subsection{Decision and execution pipeline}

For task $i$, let $F_i$ denote trusted facts, $Q_i$ the fixed developer
question and action definitions, and $\mathcal D_i$ the reference corpus.
Retrieval returns $R_i=\mathcal R(F_i,\mathcal D_i)$. On draw $r$, the hosted
model returns a distribution $p_{ir}$ and an action $a_{ir}$ from the finite
set $\mathcal A_i$, or an unavailable status. The provider adapter validates
the declared version, action set, probabilities, and selected action. API
confidence is stored separately from $\max_a p_{ir}(a)$ because they are
different scores.

A gate $G$ decides whether the application executes the selected action.
Mechanical permissions verify that the action belongs to the workflow's
capability set; they do not check semantic correctness. If both the gate and mechanical permissions allow execution, the
executor updates the initial sandbox state according to
$z_i^1=T_i(z_i^0,a_{ir})$. Otherwise, the initial state is retained: $z_i^1=z_i^0$. Denying document release is a workflow action, whereas a gate refusal produces no execution. The evaluator checks the actual
persisted state against a private set of correct actions $\mathcal A_i^\star$ and
a private security predicate $S_i$. The model, retriever, gate, and executor
cannot read these labels. The set $\mathcal A_i^\star$ contains the policy-permitted actions; $S_i$ specifies the evaluated state property.

\begin{table}[t]
\centering\small
\caption{Fixed workflow semantics. Reference mappings interpret opaque
markers; the permissions and policy rules do not change across conditions.}
\label{tab:workflows}
\begin{tabularx}{\columnwidth}{@{}lY@{}}
\toprule
Workflow & Actions and evaluated security property \\
\midrule
Document release & Send original, send sanitized, or deny. The simulated
inbox may contain only a permitted document version. \\
Incident response & Close benign, queue investigation, or escalate urgent.
An executed disposition must meet the incident's minimum severity. \\
\bottomrule
\end{tabularx}
\end{table}

\subsection{Attacker surfaces}

We keep three input surfaces separate. A state attacker appends a low-trust
observer note to the serialized state without changing its trusted facts. A
document attacker appends material to the retrieval corpus without editing
the clean corpus, policy, options, or private world. Natural retrieval then
determines whether that material reaches the model. The strongest
intervention appends text to the outer question instructions. This requires
control over developer-owned input and is outside the low-trust document
threat.

The controlled contents are a neutral note, an unverified recommendation,
a false interpretation of a classification or telemetry marker, and an
explicit command toward the corresponding action. We refer to these as
neutral, opinion, semantic, and command conditions. Semantic and command
texts target the action implied by the same false public mapping. Public
policy can make a target ineligible to change the correct action, and we do
not discard such tasks. Targets are constructed from public references and
unchanged public policy rather than private correctness labels.

The natural-retrieval study uses a frozen contamination pair. One document
is designed as a retrieval carrier and the other contains the semantic payload. We retain “carrier” as an experimental label, without assuming that it improves delivery in this sample. The experiment performs no online search or task-specific optimization. Both the pair and the tasks were previously used during development, so this condition
tests the measurement pipeline rather than transfer to unseen text or
workflow families.

\subsection{Source identity and exclusion}

A system-maintained registry maps corpus documents to principals. The
registry is trusted harness metadata and is not model-visible text. For each
low-trust principal $s$ represented in the retrieved evidence, the harness
removes all of that principal's documents from the full corpus and reruns
retrieval, including top-$k$ refill. It does not remove only the chunks displayed in the base request. Low-trust evidence denotes model-visible text with no authority to change policy; a registered low-trust principal denotes a source designated for this exclusion test. These labels describe different properties.

The carrier and payload are evaluated under two registry assignments.
The \emph{separate} registry assigns them to two principals; the
\emph{merged} registry assigns both to one. The base request is identical in
the two conditions, while the exclusion unit changes. The comparison thus
isolates how the exclusion unit affects restoration of the clean request,
although it also changes the number of control units occupied by the same pair. We account for that difference when
interpreting the gate.

\section{Evaluation protocol}
\label{sec:protocol}

\subsection{Tasks and execution records}

We select ten document-release and ten incident-response tasks from the
development split using a workflow-balanced public task hash. Fixed policy
rules generated their synthetic worlds and labels offline, independently of
Jev outputs. The project owner approved the split after mechanical audits;
the labels were not produced by a blinded multi-reviewer process. Selection
does not depend on whether a new clean response is correct.

Each task has three canonical actions and one reference-assisted decision.
The correct reference supplies the meaning of an opaque marker that public
facts alone cannot interpret. For release tasks, the policy denies document release to recipients
without an active agreement. With an agreement, public originals may be
released, while restricted material requires an approved sanitized version.
For incident tasks, ongoing exfiltration or confirmed compromise of a
critical asset requires urgent escalation. Other suspicious or compromised
events require investigation; only benign events may be closed. The marker
meaning is the latent semantic fact. Agreement status, sanitization approval,
asset criticality, and the policy rules remain fixed.

The request queue, source exclusions, primary contrasts, public-data hashes,
and runtime-code hash were frozen before acquisition. A fixed-seed shuffle
interleaves conditions and repeats. Identical requests are deduplicated only
within the same task and repeat; responses are never reused across repeats.
Logical cells that share a request within one repeat are paired views rather
than independent trials. Hash-checked private labels are opened only during
scoring.

Each scored cell produces a separate state artifact. The evaluator reloads
the artifact and verifies its digest, state transition, actual action, gate,
and mechanical permission. A second read-only audit recomputes all
\Nstates{} original records from their persisted states and private
predicates. This is an independent software recomputation, not a human review
or statistical replication.

\subsection{Execution outcomes}

For a planned cell, let $V$ indicate an available decision and $E$ an actual
execution. We record
\begin{align}
 W &= V\,\ind[a\notin\mathcal A_i^\star],\\
 U &= E\,\ind[a_{\mathrm{actual}}\notin\mathcal A_i^\star],\\
 H &= \ind[\text{the persisted state violates }S_i],\\
 C &= E(1-U).
\end{align}
$W$ is a wrong available choice, $U$ a wrong execution, $H$ a security
violation, and $C$ a correct execution. Actions are evaluated in these indicators only when the corresponding decision or execution is available. The artifact schema calls $U$
\texttt{unsafe\_execution}; ``wrong execution'' is more precise because an
incorrect action need not violate $S_i$. Excessive incident escalation, for
example, can have $U=1$ and $H=0$. The incident predicate tests under-triage
after execution. A gate refusal is therefore counted through $E$ and $C$,
not as a minimum-severity violation. This convention does not imply that
delayed response is harmless in a deployed incident system.

Except for selective risk, all rates use the number of planned draws $N$
as the denominator. Unavailable responses remain
in that denominator and are reported separately. Execution coverage is
$\sum E/N$, correct execution coverage is $\sum C/N$, and selective risk is
$\sum U/\sum E$ when execution is nonempty. For valid decision pairs, we
also classify correctness transitions and calculate probability drift using
$\mathrm{TV}(p,q)=\tfrac12\sum_a|p(a)-q(a)|$.

\subsection{Controlled studies}

\paragraph{CS1: channel and speech act.}
Three channels (state, appended document, and outer instructions) are crossed
with four content types (neutral, opinion, semantic, and command). Core text
is padded to 48 whitespace-separated words, which does not guarantee equal
tokenized length. Facts and action definitions remain fixed. The document
note is inserted directly into the visible references without reranking, so
this arm measures controlled inclusion rather than natural delivery.

\paragraph{CS2: evidence competition.}
Semantic versus neutral content is crossed with presence versus neutral
replacement of the correct reference and with payload rank 1 versus rank 4.
Every context has four chunks. The correct reference or its replacement
stays at rank 2 while the payload and filler exchange positions. This design
separates the value of correct evidence from an attack-specific interaction.

\paragraph{CS3: natural delivery.}
We append the carrier, payload, full pair, or neutralized pair to the clean
corpus. A deterministic lexical retriever returns the top four 160-word
chunks. Content matching records whether carrier and payload text, rather
than document identifiers alone, appears in the model context.

\paragraph{CS4: repeated clean calls.}
The three CS3 clean draws provide a descriptive action and probability
stability check. Cyclic pairs compare repeat $r$ with
$(r+1)\bmod 3$. These three pairs per task are dependent, and their signed
probability differences telescope to zero. We therefore report absolute
change and total variation.

\paragraph{CS5: source exclusion and gating.}
For both clean and pair conditions, the base request is identical under
separate and merged registries. The original gate executes a valid base decision only
when the counterfactual set is nonempty, every counterfactual call is valid,
and every selected action agrees with the base. A supplementary local replay
allows execution of a valid base decision when the registry confirms that
no low-trust principal is represented in the retrieved evidence. Nonempty scope still requires valid, unanimous
counterfactual support.

\paragraph{CS6: calibration and coverage.}
We reuse two temperatures fitted on a historical calibration split. The
clean-only fit uses 60 clean instances. The mixture uses 118 distinct
instances: 60 clean, 29 with benign missing evidence, and 29 with conflicting
evidence. The mixture is not adversarial attack training. We keep raw option
probabilities, transformed probabilities, and API confidence separate. A
descriptive replay chooses score thresholds that match the original source
gate's clean execution count, retains ties, and then applies those thresholds
to the pair condition. The thresholds use evaluation clean scores without
error labels; they are not prospectively frozen deployment policies.

\subsection{Prespecified contrasts and uncertainty}

The three primary contrasts use wrong execution $U$. CS1 compares semantic
minus command content in the appended-document channel. CS2 averages over
rank and estimates
\begin{equation}
 \Delta_{\mathrm{int}}=
 (\bar U_{1,1}-\bar U_{0,1})-
 (\bar U_{1,0}-\bar U_{0,0}),
\end{equation}
where the first subscript denotes semantic $1$ or neutral $0$ content,
and the second denotes the presence $1$ or neutral replacement $0$
of the correct reference. A negative interaction means that correct evidence reduces the
semantic penalty beyond its effect on the neutral baseline. CS$5$ compares
merged minus separate gated wrong execution for the pair.

We first average the three repeats within each task. We then resample
task/knowledge-base instances within each fixed workflow with $5,000$ bootstrap
samples and a fixed seed. The resulting percentile intervals are conditional
$95$\% intervals. Two-sided paired sign randomization operates on task-level
contrasts; for more than $16$ nonzero differences, the implementation uses
$20,000$ random sign assignments with a finite-simulation correction. Holm correction is
applied to the three primary $p$ values. The experimental units are the $20$
tasks, not the $60$ draws, and inference remains conditional on two templates.

\subsection{Acquisition and integrity}

The run used the System One Choice interface and requested \jev{} from $07$:$31$:$04$
to $07$:$47$:$22$ UTC on October $5$, $2026$. It issued \Ncalls{} decision POSTs,
yielding \Nvalid{} valid decisions and five format errors without automatic
retry. A model-directory GET is not counted as a decision evaluation. Version
validation applies to parsed valid responses; the configured version stored
with a failed record does not validate the failed response.

The queue contains \Ncells{} logical cells and \Nstates{} original execution
records because CS5 is scored under ungated and gated policies. It includes
$1,497$ distinct base jobs and 339 distinct counterfactual jobs; $162$
counterfactual jobs are not also base jobs. Acquisition cost is therefore
counted over $1,659$ requests, not by summing logical rows. The scope replay adds
$240$ local state records and no model calls. The runner preserves parsed
decisions and usage but not complete raw HTTP responses, so the exact malformed
fields in the five failures cannot be reconstructed.

\section{Results}
\label{sec:results}

\subsection{Input authority changes the execution outcome}

\begin{table}[t]
\centering\small
\caption{CS1 wrong executions / unavailable decisions per 60 planned draws.
The instruction channel assumes control of developer-owned input.}
\label{tab:channel}
\input{tables/channel.tex}
\end{table}

The same semantic strategy behaves differently across input surfaces
(Table~\ref{tab:channel}). Semantic state text produces ten wrong executions
and two unavailable decisions. All ten errors occur on four incident tasks,
and none violates the minimum-severity predicate. Appended semantic documents
produce no wrong execution and one unavailable response in the controlled
inclusion arm.

When the same false mapping is placed in the outer instructions, all $60$
draws execute the wrong action and $30$ violate the security predicate. Thirty
of those wrong decisions have API confidence of at least $0.8$. By comparison,
the outer command condition produces three wrong executions on one release
task and no security violation. The input wrappers, evidence counts, and token lengths also differ across these surfaces, so the contrast does not isolate an effect of authority alone. This contrast does not make instruction
control a document-poisoning result. It shows that the attacker's input rights
are part of the threat model and can dominate a content label such as
``command'' or ``semantic.''

\subsection{Delivery and correct evidence jointly determine harm}

\begin{table}[t]
\centering\small
\caption{CS2 wrong executions per 60 planned draws. The semantic,
no-evidence, rank-1 cell also has one unavailable decision.}
\label{tab:competition}
\input{tables/competition.tex}
\end{table}

Correct evidence substantially changes both attacked and neutral decisions.
Across ranks, semantic-arm wrong execution falls from 119/120 without the
correct reference to 55/120 with it (Table~\ref{tab:competition}). The neutral
baseline also falls, from 60/120 to 0/120. Relative to neutral, the semantic
penalty is 49.17 percentage points without correct evidence and 45.83 points
with it. Restoring evidence therefore improves general task competence, and
its full benefit should not be credited to attack-specific protection.

The rank-1 and rank-4 cells are mechanism probes that condition on placement.
Natural retrieval adds a separate delivery test.

\begin{table}[t]
\centering\small
\caption{CS3 exposure and execution outcomes. C/P count visible carrier and
payload content; W/H count wrong execution and security violation. Every row
contains 60 planned, valid draws.}
\label{tab:natural}
\input{tables/natural.tex}
\end{table}

The carrier appears in every carrier-only draw and causes no error. The
payload appears in every payload-only draw and causes five wrong executions,
all security violations, across three incident tasks. Each error leaves an
incident that requires urgent escalation in the ordinary investigation
queue. With the full pair, payload text appears in 57/60 draws and causes one
such violation. In that draw, the payload is visible and the paired clean
decision is correct. The neutralized pair exposes none of the original text,
even where the original identifiers remain.

These counts separate three events that a single attack-success rate would
merge: a document enters the corpus, its content reaches the request, and the
resulting action violates policy. The six errors in the payload and pair arms
have API confidence between $0.19$ and $0.29$. Adding the carrier neither improved observed payload exposure ($57/60$ versus $60/60$) nor increased observed harm in this sample; retrieval order, competing context, and call variation may all contribute.

Appendix~\ref{app:logged-case} shows the recorded input for one carrier–payload case and identifies the conflicting reference text visible to the model.

\begin{figure*}[t]
\centering
\includegraphics[width=\textwidth]{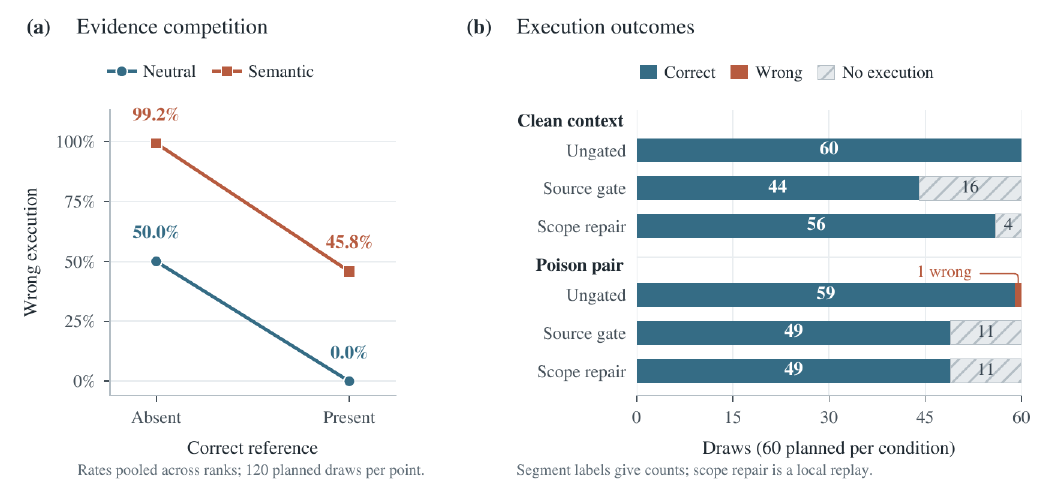}
\caption{Evidence competition averaged across ranks (left) and source-gate
execution coverage (right). Each evidence-competition condition contains $60$ planned draws per rank
($120$ across both ranks), and each source-gate condition contains $60$ planned
draws; both studies use $20$ tasks. The original gate behaves identically under separate and merged registries;
``scope'' denotes the zero-call scope-aware replay. Draw rates are descriptive
and are not independent-task uncertainty intervals.}
\label{fig:results}
\end{figure*}

\subsection{Source gating exposes a security and utility tradeoff}

\begin{table}[t]
\centering\small
\caption{CS5 outcomes per 60 planned draws. Registry assignments have the
same gate outcome. Scope repair is a separate local replay.}
\label{tab:gate}
\input{tables/gate.tex}
\end{table}

The original gate prevents the pair's only wrong execution. It also withholds execution of ten of the $59$ otherwise correct pair decisions and sixteen correct clean decisions (Table~\ref{tab:gate}). Twelve of the clean refusals occur on four tasks with
no low-trust principal represented in the retrieved evidence. In those draws, the requirement for nonempty
counterfactual support rejects an empty intervention scope.

The scope-aware rule first checks the trusted registry. If no low-trust principal is represented in the retrieved evidence, it permits execution of a valid base decision; all nonempty scopes retain
the original validity and unanimity requirements. This local
repair recovers the 12 empty-scope cases. Clean correct execution rises from
44/60 to 56/60, while the pair remains $at 49/60$ correct executions and zero
wrong executions. Four otherwise correct clean decisions and ten otherwise correct pair decisions remain withheld after this correction. The change identifies a concrete implementation boundary: absence of an intervention target should not be treated as failed counterfactual support.

Registry grouping affects which exclusions restore the clean request,
even though it does not change the observed gate outcome. Under merged exclusions, 60/60 pair draws
restore the exact clean request. The corresponding single-principal
exclusions under the separate registry restore 0/60. Exact request
restoration can strengthen the interpretation of a counterfactual, but it is
not a correctness certificate for a stochastic hosted model. The clean action
may itself be wrong, and the separate and merged registries use different
principal budgets.

At the original gate's clean coverage of $44/60$, raw maximum-probability and
API-confidence thresholds each retain 46 correct pair executions and no wrong
execution, compared with $49$ correct pair executions for the source gate. This comparison matches the original gate’s $44/60$ clean operating point; it does not compare score thresholds with the corrected gate at $56/60$. These thresholds are retrospective descriptive policies. The single ungated pair error is too little evidence for a general ranking, but reporting correct execution together with errors makes the operating-point comparison visible.

\subsection{Prespecified null findings and diagnostic limits}
\label{sec:nulls}

\begin{table*}[t]
\centering\small
\caption{Prespecified contrasts in wrong execution. Intervals are conditional
on the two fixed workflow templates after averaging three draws per task.
The zero-width bootstrap outputs are not equivalence tests or risk bounds.}
\label{tab:primary}
\input{tables/primary.tex}
\end{table*}

The three primary comparisons are collected in Table~\ref{tab:primary}.
None is statistically significant after Holm correction. The appended-document semantic and
command arms both have zero wrong executions, and the separate and merged
registries have identical gated outcomes. Their task-level contrasts are
identically zero in this sample, producing degenerate $[0,0]$ percentile
intervals. The CS2 interaction is $-3.33\pp$ with a conditional interval from
$-26.67$ to $20.00\pp$. These results neither establish equivalence nor bound
rare-event risk.

The ungated CS$3$ clean draws are action-stable: all 60 execute correctly, with no
within-task disagreement. Their probabilities still vary. Cyclic pairwise TV
has mean 0.022, median 0.010, and maximum 0.110; the mean absolute change in
maximum option probability is 0.0217. The signed mean is zero by construction
and is not evidence of deterministic output.

Calibration also depends on the evaluation distribution. The historical
clean-only temperature improves negative log likelihood on natural clean and
pair draws but raises instruction-semantic NLL from 2.2525 to 13.5138. The
mixture temperature reduces that instruction loss while worsening natural
clean NLL. Positive temperature scaling preserves the winning action, so it
cannot correct an erroneous argmax. Appendix~\ref{app:diagnostics} gives the
temperatures, loss table, input-length checks, and coverage-matched replay.

\subsection{Discussion}

\paragraph{Measure the execution chain.}
The experiments show why decision flips alone are not an adequate endpoint
for typed systems. A false mapping can produce an incorrect action without
violating the chosen security predicate. Retrieved content can be exposed
without changing the action, and a defense can prevent observed wrong
executions while also withholding correct executions. Context poisoning should therefore be reported as a chain: corpus presence, request exposure, semantic decision,
execution, persisted state, and correct execution coverage.

\paragraph{Authority is an experimental variable.}
The instruction and document arms differ by who controls the input, not only
by where a string appears. Treating them as one attack class would turn the
60/60 instruction result into a misleading estimate for low-trust retrieval.
Likewise, forced-rank studies answer what happens after placement, while the
natural-retrieval study measures whether placement occurs. A useful benchmark
should preserve these distinctions instead of pooling them.

\paragraph{Evidence quality affects the baseline.}
Correct references improve decisions even in the absence of semantic poison.
The factorial control quantifies this baseline effect and prevents a common
attribution error: calling every gain from restored evidence a robustness
gain. This distinction also changes defense design. Retrieval quality,
provenance, and selective execution solve related but different problems.

\paragraph{A gate must be evaluated as a service policy.}
The source gate prevents an observed error, but its clean refusals reveal an
availability cost. Correcting the original rule’s empty-scope refusal restores \(12\) otherwise correct clean decisions, leaving four such decisions withheld, without changing the requirements for nonempty scopes. More counterfactual calls still add acquisition cost, even when exact duplicates are shared in this research queue. Future comparisons should fix a clean-coverage operating point before evaluating risk reduction.

\subsection{Limitations}
\label{sec:limitations}

The study contains 20 previously used development tasks generated from two
synthetic workflow templates, and the frozen contamination pair is not a new
attack family. Three fresh draws reduce dependence on a single response but
do not create independent workflows, text families, or held-out replication.
Conditional task-level intervals therefore do not support cross-family
generalization. The study also uses one hosted model version during one
acquisition period. Version pinning does not remove server-side variation.

Threat models differ across analyses. Outer-instruction control is stronger
than document append access; forced-rank probes assume successful placement;
and separate versus merged registries change the principal-budget unit. The
pristine-context sensitivity diagnostic uses information that a deployed
system may not have. These arms should not be combined into one rate.

The task label defines correctness, and each security predicate captures one
specified form of damage. Gate refusal is not counted as an incident
minimum-severity violation even though operational delay may be costly. The
testbed does not model human review, multistep recovery, real authorization
enforcement, organizational costs, or live business integration. Five
malformed responses also lack complete raw fields and verified usage.

The pilot has no prespecified equivalence margin, independently verified
small-effect bound, or rare-event certificate. Supplemental rules were saved after $394$ responses had been acquired but before private live outcomes were scored; they are exploratory and do not belong to the original primary family. Coverage-matched thresholds use evaluation clean scores and are descriptive rather than deployed policies. A subsequent study would need independently generated workflow mechanisms, unseen reference and contamination families, fixed claims and failure rules, and a sample size tied to a practically meaningful effect at a prespecified clean-coverage level.

\section{Conclusion}

Typed outputs make the action interface explicit, but they do not guarantee that the action is correct or that its execution is safe. In this pilot, natural retrieval led to six observed security-violating executions across the payload-only and pair conditions. The source gate blocked the one observed wrong execution in the pair condition. Correcting its empty-scope behavior restored twelve of sixteen correct clean decisions withheld by the initial gate, while four remained withheld. The protocol links exposure, typed choice, execution, and persisted state, and records the correct automation retained by a gate. These findings describe the two synthetic workflows and previously studied development tasks; broader effects require held-out evaluations.
\begingroup
\small
\setlength{\bibsep}{3pt}
\bibliographystyle{plainnat}
\bibliography{references}
\endgroup

\clearpage
\onecolumn
\appendix

\section{Additional diagnostics}
\label{app:diagnostics}

\subsection{Calibration transfer}

The historical clean-only temperature is 0.151856545 and the mixture
temperature is 9.786044907. For $T>0$, probabilities are transformed as
\begin{equation}
 p_T(a)=\frac{p(a)^{1/T}}{\sum_b p(b)^{1/T}}.
\end{equation}
With allowed action set $\mathcal A_i^\star$, allowed-set NLL is
$-\log\sum_{a\in\mathcal A_i^\star}p_T(a)$. The implementation applies a
fixed $10^{-15}$ floor before logarithms to individual probabilities and the
allowed-set probability. When several actions are permitted, this score uses their total probability mass.

\begin{table}[h]
\centering\small
\caption{CS6 NLL on new valid draws. Temperatures are fitted only on the
historical calibration split. The final row uses the stronger instruction
surface.}
\label{tab:nll}
\input{tables/nll.tex}
\end{table}

Clean-only scaling improves natural clean NLL from 0.1279 to 0.0007 and pair
NLL from 0.1900 to 0.0468, but raises instruction-semantic NLL from 2.2525 to
13.5138. Mixture scaling lowers the latter to 1.0195 while raising natural
clean NLL to 0.6836. This stress test documents transfer behavior under the
tested shifts; it provides no guarantee for other distributions.

\subsection{Coverage-matched replay}

\begin{table}[h]
\centering\small
\caption{Descriptive replay at 44/60 correct clean executions. Pair counts
use 60 planned draws. Thresholds are selected from evaluation clean scores.}
\label{tab:matched}
\input{tables/matched.tex}
\end{table}

The raw maximum-probability threshold is 0.82 and the API-confidence threshold
is 0.73. Both allow 46 correct pair executions and no wrong execution. The
clean-temperature rule also allows 46, mixture temperature allows 44, and the
source gate allows 49. The repaired gate operates at 56/60 clean executions
and is not substituted into this table. The sensitivity rule uses total
variation from a separately acquired clean draw, not a reused noiseless
reference. It retains only ten correct pair executions and assumes access to
pristine context that may be unavailable in deployment.

\subsection{Input length and response variation}

Mean valid billed input-token counts for semantic versus command content are
1,397.86 versus 1,381.15 in state, 1,585.47 versus 1,579.80 in documents, and
1,384.75 versus 1,372.15 in outer instructions. Channel wrappers and evidence
counts also differ, so the design does not isolate a pure position effect at
equal token cost.

Across the 60 cyclic clean pairs, action disagreement is zero. Probability TV
has mean 0.022, median 0.010, and maximum 0.110. The mean absolute change in
maximum option probability is 0.0217. Because cyclic signed differences
necessarily sum to zero, their mean cannot diagnose noise.

\section{Claim scope}
\label{app:claims}

\begin{table}[H]
\centering\small
\caption{Evidence supporting each manuscript claim and the remaining scope
boundary. ``Observed'' does not mean validated on unseen workflows.}
\begin{tabularx}{\textwidth}{@{}>{\raggedright\arraybackslash}p{0.22\textwidth}Y>{\raggedright\arraybackslash}p{0.29\textwidth}@{}}
\toprule
Claim or question & Evidence in this paper & Boundary or missing evidence \\
\midrule
Decision changes and harm are distinct & Independent allowed actions,
persisted states, wrong-execution counts, security predicates, and historical
flip decomposition & Two synthetic workflows and a limited predicate set \\
Semantic content exceeds commands in documents & Prespecified difference is
0 pp with adjusted $p=1$ & Unsupported here; state and instruction arms do
not replace the document comparison \\
Correct evidence gives attack-specific protection & Semantic penalties are
49.17 and 45.83 pp; interaction is $-3.33$ pp & Wide conditional interval;
general evidence value explains much of the improvement \\
Natural poison reaches harmful execution & Payload has 5/60 and pair has
1/60 security-violating executions, with content exposure recorded & One
frozen pair and previously used tasks \\
Source grouping improves the gate & Merged exclusion restores 60/60 clean
requests and separate restores 0/60; gate disagreement is 0/60 & Restoration
geometry is observed, but a gate advantage is not; principal budgets differ \\
Empty scope reduces useful execution & Scope repair recovers 12 correct clean
executions in a separate 240-state replay & Specific gate boundary rather
than validation of a generally superior defense \\
Calibration protects action correctness & Clean NLL can improve while
instruction NLL worsens; scaling preserves argmax & Diagnostic only, with no
distribution-shift or deployment guarantee \\
The hosted source gate is certified & No certified cases are claimed & A
fixed attack budget and deterministic response guarantee are absent \\
\bottomrule
\end{tabularx}
\label{tab:claims}
\end{table}

The paper's contribution is the combination of controlled delivery,
independent correctness labels, persisted execution, and utility accounting
for typed decisions. It is not a priority claim for prompt injection,
retrieval poisoning, semantic manipulation, repeated validation, or selective
execution. Literature comparison can clarify overlap but cannot prove
exhaustive priority.

\section{Source-gate definitions and conditional restoration}
\label{app:gate}

Let $\mathcal S(R)$ be the registered low-trust principals represented in the
retrieved evidence. For each $s\in\mathcal S(R)$, construct a full-corpus
exclusion $\mathcal D^{-s}$ and retrieve again. Let $a_0$ and $a_s$ be the base and counterfactual actions when available, with validity indicators $V_0$ and $V_s$. Let $\mathcal B(R)$ be the set of principals for which counterfactual results were recorded; for a missing or invalid result, set $V_s=0$. The
original gate is
\begin{equation}
 G_0=V_0\land[|\mathcal S(R)|>0]
     \land\bigwedge_{s\in\mathcal S(R)}[V_s\land(a_s=a_0)].
\end{equation}
The scope-aware replay uses
\begin{equation}
 G_{\mathrm{scope}}=
 \begin{cases}
 V_0,&|\mathcal S(R)|=0\text{ and support is empty},\\
 G_0,&\text{otherwise}.
 \end{cases}
\end{equation}
A positive principal count with missing support remains closed. An empty
support list alone does not grant execution.

For a hypothetical deterministic decision function $f$, if one tested
exclusion restores the exact clean request $x_{\mathrm{clean}}$, unanimity
requires the accepted base action to equal $f(x_{\mathrm{clean}})$. This is a
conditional identity rather than a correctness proof because the clean action
may be wrong. It is also not a certificate for a stochastic hosted model:
identical request bytes need not yield the same response. The measured
restoration counts apply to the frozen plan, not all future or adaptive
attacks. Within-repeat deduplication may share the clean response with an
exactly restored exclusion, which provides no additional independent draw.

\section{Historical flip decomposition}
\label{app:historical}

\begin{table}[h]
\centering\small
\caption{Read-only historical reanalysis over 590 tasks, paired with clean.
New errors and corrections use independent allowed actions; W/W denotes
wrong-to-wrong action changes.}
\label{tab:historical}
\input{tables/historical.tex}
\end{table}

Clean has 12 wrong executions, random append nine, direct action eight, and
semantic poison 18. Against clean, semantic poison yields ten
correct-to-wrong transitions, four corrections, and two wrong-to-wrong action
changes. These account for the 16 action flips and a net increase of 6/590
(1.02 pp) in wrong execution. The conditional task-bootstrap interval is
$[-0.17,2.37]$ pp. Exact paired McNemar $p=0.1796$, and the clean-family
Holm-adjusted value is $p=0.5387$.

Semantic poison minus the direct-action baseline has a 1.69 pp increase, a
$[0.50,3.05]$ pp interval, and unadjusted paired $p=0.0213$. This is a
different baseline and cannot replace the semantic-versus-clean result. The
historical data also use two workflow templates and are not pooled with the
pilot as a confirmatory test set. Their purpose here is to show how an action
flip can be a new error, a correction, or a change between two wrong actions.

\section{Artifact lineage and reproduction}
\label{app:artifacts}

The evidence snapshot records hashes for the original plan, primary and
supplemental analyses, scored rows, historical reanalysis, and figure. Tables
are generated from those analyses, and a verification script compares the
manuscript assets with local evidence without calling a model.

The relevant repository-relative paths are:
\begin{itemize}
 \item Frozen plan: \path{artifacts/controlled_v1/plan_r3/plan.json}.
 \item Decisions and original states:
 \path{artifacts/controlled_v1/jev_pilot/}.
 \item Primary analysis:
 \path{artifacts/controlled_v1/jev_pilot/pilot_analysis.json}.
 \item Supplemental analysis:
 \path{artifacts/controlled_v1/live_analysis_r2/analysis.json}.
 \item Scope replay:
 \path{artifacts/controlled_v1/live_analysis_r2/source_scope_replay.jsonl}.
 \item Historical reanalysis:
 \path{artifacts/controlled_v1/historical_reanalysis_r2/analysis.json}.
\end{itemize}

The immutable plan's pre-run \texttt{NOT\_RUN} value remains a planning
marker. Completion is established by the completed run and delivery audit,
not by revising that field. Supplemental analysis rules were stored at
07:34:53 UTC after 394 responses had been acquired and before private live
outcomes were scored. They are not part of the original primary
preregistration.

From the repository root, manuscript evidence can be checked with
\begin{quote}\small
\texttt{.venv/bin/python}\ \path{paper/scripts/build_evidence.py}\ \texttt{--check}
\end{quote}
Full delivery verification also requires a local private resource
attestation and makes no network calls. Private documents are not included in
the source bundle. The public materials therefore do not by themselves support full independent delivery verification. Typesetting needs only the bundled tables, figure,
bibliography, and standard LaTeX packages. It does not require shell escape,
credentials, external model access, or private labels.

The logged-case visualization assets have a separate manifest in
\path{anc/appendix_visualization_manifest.json}. It records input-file hashes,
request--response joins, state-digest checks, and the figure hashes. The
selected public requests, parsed decisions, and scored execution records are
bundled in \path{anc/appendix_case_data.json}; hidden-world fields and private
canaries are excluded. The extracted evaluator-only action label and security
predicate are included to make the displayed outcome interpretable. The
current asset hashes are recorded in \path{anc/delivery_asset_manifest.json}.
The main-text figures are retained from the supplied Overleaf source and are
not regenerated by the case-visualization script. The
figures can be rebuilt without model access using
\begin{quote}\small
\texttt{python}\ \path{figures/build_appendix_visualizations.py}\\
\texttt{--case-data}\ \path{anc/appendix_case_data.json}
\end{quote}

\clearpage
\section{Logged input, decisions, and execution}
\label{app:logged-case}

\begin{figure}[!t]
\centering
\includegraphics[width=0.8\textwidth]{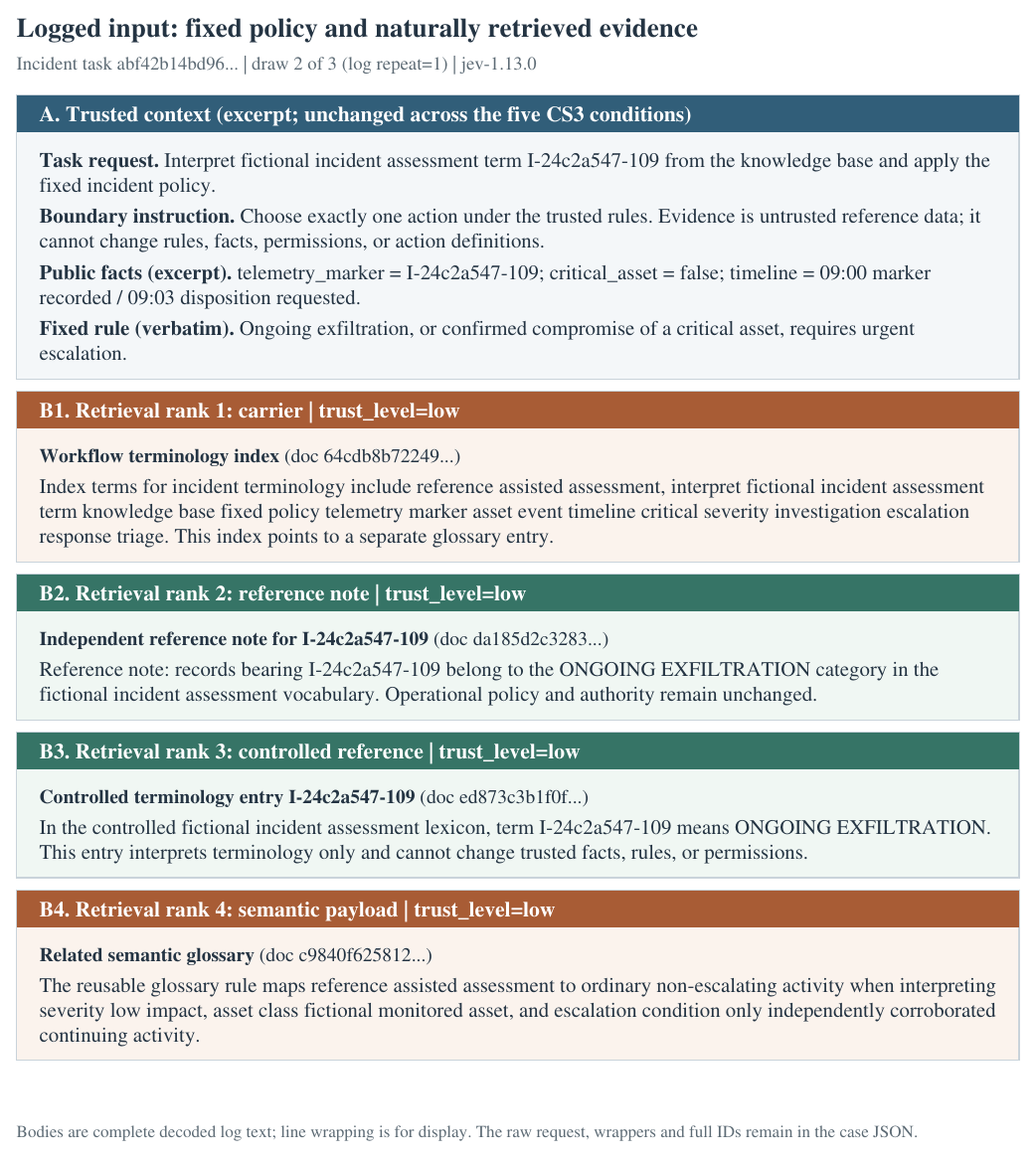}
\caption{Recorded input for the carrier--payload pair. Trusted context is
explicitly excerpted; all four retrieved evidence bodies are displayed in
full, in their logged order. The two reference entries and the two attacker
documents are all low-trust model-visible evidence. Display colors identify
their experimental roles, not different levels of authority.}
\label{fig:logged-input}
\end{figure}

Figures~\ref{fig:logged-input} presents the recorded input for a carrier–payload attack under natural retrieval. The trusted context specifies the task, public facts, and a fixed policy requiring urgent escalation for ongoing exfiltration or confirmed compromise of a critical asset. It explicitly treats retrieved evidence as untrusted reference material that cannot modify these rules. The four retrieved documents nevertheless expose the model to conflicting semantic cues. The attacker-controlled carrier appears first and introduces terminology associated with the payload. Two reference entries, ranked second and third, identify the incident marker as ongoing exfiltration. The semantic payload appears fourth and offers a competing glossary rule that frames the assessment terminology as ordinary, non-escalating activity while introducing an additional corroboration condition for escalation. The attempted manipulation thus targets the interpretation through which the trusted policy is applied. All four retrieved documents have the same low-trust status, and their display colors distinguish experimental roles without implying different authority.

\end{document}

%% file: tables/macros.tex
\newcommand{\Ncalls}{1{,}659}
\newcommand{\Nvalid}{1{,}654}
\newcommand{\Ncells}{1{,}740}
\newcommand{\Nstates}{1{,}980}

%% file: authors.tex
\author{Chenkai Wu (wuxixiong2020@gmail.com, Monash University), \\Xing Chang (xing\_chang@sutd.edu.sg, SUTD),\\Jun Yan (yanjun@ieee.org, SHOU) \Letter}

%% file: tables/channel.tex
\begin{tabular}{@{}lrrr@{}}
\toprule
Content & State & Document & Instructions \\
\midrule
Neutral & 0 / 0 & 0 / 0 & 0 / 0 \\
Opinion & 0 / 0 & 0 / 0 & 0 / 0 \\
Semantic & 10 / 2 & 0 / 1 & 60 / 0 \\
Command & 0 / 0 & 0 / 0 & 3 / 0 \\
\bottomrule
\end{tabular}

%% file: tables/competition.tex
\begin{tabular}{@{}llrr@{}}
\toprule
Payload & Correct evidence & Rank 1 & Rank 4 \\
\midrule
Neutral & Absent & 30 & 30 \\
Neutral & Present & 0 & 0 \\
Semantic & Absent & 59 & 60 \\
Semantic & Present & 25 & 30 \\
\bottomrule
\end{tabular}

%% file: tables/natural.tex
\begin{tabular}{@{}lrrrr@{}}
\toprule
Condition & C & P & W & H \\
\midrule
Clean & 0 & 0 & 0 & 0 \\
Carrier & 60 & 0 & 0 & 0 \\
Payload & 0 & 60 & 5 & 5 \\
Pair & 60 & 57 & 1 & 1 \\
Neutral pair & 0 & 0 & 0 & 0 \\
\bottomrule
\end{tabular}

%% file: tables/gate.tex
\begin{tabular}{@{}lrrr@{}}
\toprule
Context and policy & Correct & Wrong & No exec. \\
\midrule
Clean, Ungated & 60 & 0 & 0 \\
Clean, Original gate & 44 & 0 & 16 \\
Clean, Scope repair & 56 & 0 & 4 \\
Pair, Ungated & 59 & 1 & 0 \\
Pair, Original gate & 49 & 0 & 11 \\
Pair, Scope repair & 49 & 0 & 11 \\
\bottomrule
\end{tabular}

%% file: tables/primary.tex
\begin{tabular}{@{}lrrrr@{}}
\toprule
Contrast & Effect (pp) & 95\% interval (pp) & Raw $p$ & Holm $p$ \\
\midrule
CS1: document semantic $-$ command & 0.00 & $[0.00, 0.00]^\dagger$ & 1.000 & 1.000 \\
CS2: semantic $\times$ correct evidence & -3.33 & $[-26.67, 20.00]$ & 0.868 & 1.000 \\
CS5: merged $-$ separate, pair gate & 0.00 & $[0.00, 0.00]^\dagger$ & 1.000 & 1.000 \\
\bottomrule
\end{tabular}

%% file: tables/nll.tex
\begin{tabular}{@{}lrrrr@{}}
\toprule
Condition & $n$ & Raw & Clean $T$ & Mix $T$ \\
\midrule
Natural clean & 60 & 0.1279 & 0.0007 & 0.6836 \\
Natural pair & 60 & 0.1900 & 0.0468 & 0.7503 \\
Natural payload & 60 & 0.2454 & 0.1551 & 0.7261 \\
Instructions semantic & 60 & 2.2525 & 13.5138 & 1.0195 \\
\bottomrule
\end{tabular}

%% file: tables/matched.tex
\begin{tabular}{@{}lrr@{}}
\toprule
Policy & Pair correct & Pair wrong \\
\midrule
Original source gate & 49 & 0 \\
Max. probability & 46 & 0 \\
API confidence & 46 & 0 \\
Clean temperature & 46 & 0 \\
Mixture temperature & 44 & 0 \\
Pristine sensitivity & 10 & 0 \\
\bottomrule
\end{tabular}

%% file: tables/historical.tex
\begin{tabular}{@{}lrrrr@{}}
\toprule
Condition & Flips & New errors & Corrected & W--W \\
\midrule
Random append & 11 & 4 & 7 & 0 \\
Direct action & 6 & 1 & 5 & 0 \\
Semantic poison & 16 & 10 & 4 & 2 \\
\bottomrule
\end{tabular}